\documentclass[11pt]{article}
\usepackage{xspace}
\usepackage{graphicx}
\usepackage{amsmath}
\usepackage{amssymb}
\usepackage{color}

\definecolor{Red}{rgb}{1,0,0}
\definecolor{Green}{rgb}{0,1,0}
\definecolor{Blue}{rgb}{0,0,1}
\definecolor{Black}{rgb}{0,0,0}

\def\beq{\begin{equation}}
\def\eeq#1{\label{#1}\end{equation}}
\def\eeqn{\end{equation}}

\def\beqa{\begin{eqnarray}}
\def\eeqa#1{\label{#1}\end{eqnarray}}
\def\eeqan{\end{eqnarray}}

\let\bar=\overbar

\def\Dslash{\not{\hbox{\kern-4pt $D$}}}
\def\dslash{\not{\hbox{\kern-2pt $\del$}}}

\def\msb{{\bar{\ssstyle M \kern -1pt S}}}

\def\sin22th{\textrm{sin}^2 2 \theta}
\def\dm2{\Delta m^2}
\def\Title#1{\begin{center} {\Large {\bf #1} } \end{center}}

\begin{document}

\Title{Triggering and data acquisition for the Hyper-Kamiokande experiment}

\bigskip\bigskip


\begin{raggedright}  

{\it Debra Dewhurst on behalf of the Hyper-Kamiokande UK DAQ Group\index{Dewhurst, D.},\\
Department of Physics\\
University of Oxford\\
OX1 3RH Oxford, UK}\\

\end{raggedright}
\vspace{1.cm}

{\small
\begin{flushleft}
\emph{To appear in the proceedings of the Prospects in Neutrino Physics Conference, 15 -- 17 December, 2014, held at Queen Mary University of London, UK.}
\end{flushleft}
}

\section{Introduction}

Hyper-Kamiokande (HK) has been proposed as a next generation neutrino oscillation experiment,
capable of observing accelerator, atmospheric, solar and astrophysical neutrinos in addition to possible 
proton decays, providing a rich scientific 
program~\cite{Abe:2011ts}. To be sensitive to such physics, HK will need a robust trigger and data acquisition system (DAQ). 
Several UK institutions are participating in the development of the trigger and DAQ. Here we 
present some of the on going studies from the UK DAQ group. 

\section{Hyper-Kamiokande}

HK is designed to be the third generation underground water Cherenkov detector in Kamioka, Japan.
It will have a total (fiducial) mass of 0.99 (0.56) million metric tonnes making it 20 (25) times
larger than the Super-Kamiokande (SK) water Cherenkov detector. In the baseline design, the
HK detector is composed of two tanks with an egg-shaped cross-section 48~m wide, 54~m tall and 
250~m long. Each tank will be optically separated into five compartments to allow triggering
and event reconstruction to be performed on a per compartment basis. The baseline design
specifies an inner detector photocoverage of $20\%$ provided by 99,000 20-inch diameter 
photomultiplier tubes (PMTs). 25,000 8-inch PMTs are proposed for the outer region of the 
detector.

For long baseline oscillation studies it is planned to use the accelerator complex at J-PARC, currently used
in the T2K experiment~\cite{Abe:2011ks}. Upgrades are planned to increase the beam power to 750~kW. The beam
will pass through the T2K near detector complex located 280~m downstream from the graphite target, where upgrades are also being planned.
It will then pass through an intermediate water Cherenkov detector around 2~km from the target and then HK, 295~km 
from the target. There are currently two proposals for the intermediate detector, one of which is a gadolinium-doped water Cherenkov detector called TITUS. 

\section{Current baseline design of the DAQ and strategy}

The DAQ will collect the raw (digitized) data output from the detector electronics and write the
formatted data to storage for offline analysis. It therefore must be capable of accessing all of 
the physics of interest, whilst discarding non-physics events. The data rate is expected to be 
dominated by PMT dark noise. For PMTs with 4~kHz dark noise the estimated data rate, assuming 12
bytes per PMT hit and 100,000 PMTs, is 4.8 GB/s. The expected data rate from radioactive 
backgrounds ($^{238}\textrm{U}$, $^{232}\textrm{Th}$ and $^{222}\textrm{Rn}$) is on 
the order of 30-250 kB/s.

The current designs for triggering beam events are based on the same concept as used during Phase 4 of the SK electronics~\cite{Yamada:2010zzc}. All the 
PMT signals above a certain threshold will be digitized and read out 
as digitized hits. A simple Nhits threshold cut can then be applied whereby if the number of PMTs fired is greater than the 
Nhits threshold the event will be
read out and stored to disk. This type of threshold cut will not be suitable for low energy events ($< 100~\textrm{MeV}$). 
The UK DAQ group are currently performing studies to determine what types
of interactions we can expect to recover with the simple Nhits trigger and what interactions may be lost
and how we might recover these lost interactions. These studies will inform the technical
design of the DAQ and trigger.

\section{On going physics studies}

Studies are being performed using the official HK detector simulation package. This is a GEANT4-based~\cite{Agostinelli:2002hh} water Cherenkov
detector simulation package known as WCSim~\cite{WCSim:2015}. 
The studies
are currently focussing on developing methods to discard non-physics events and
understanding the impact of the PMT photo-coverage for being able to access low
energy events. 

\subsection{Threshold studies}

\begin{figure}[!ht]
\begin{center}
\includegraphics[width=0.45\columnwidth]{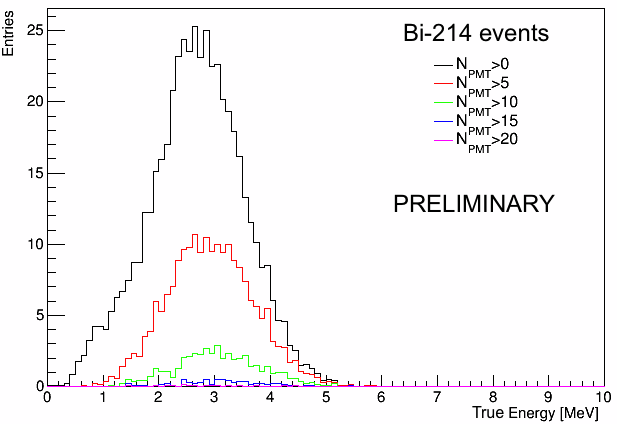}
\includegraphics[width=0.45\columnwidth]{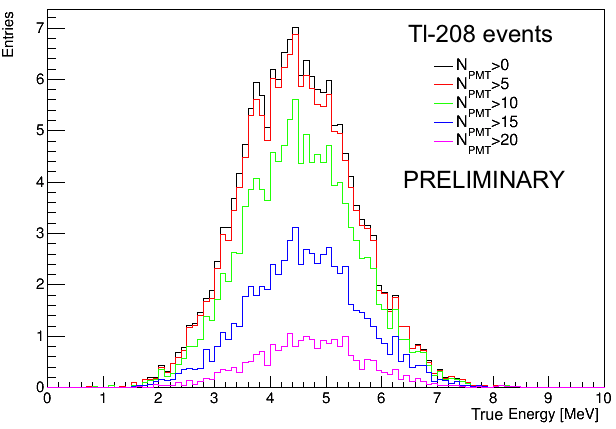}
\caption{The expected energy spectra for the radioactive decays of $^{214}\textrm{Bi}$ (left) and $^{208}\textrm{Tl}$ (right) 
as a function of the true energy, smeared by $21\%$ to account for detector energy resolution, for different Nhits thresholds.}
\label{fig:threshold}
\end{center}
\end{figure}

Studies are being performed to determine an appropriate Nhits threshold for the main HK trigger. 
Figure~\ref{fig:threshold} shows
$^{214}\textrm{Bi}$ and $^{208}\textrm{Tl}$ simulated and processed through WCSim v1.1.1 to look at the effect of the Nhits 
threshold on the true energy spectrum for low energy events. Similar studies are being performed 
for other physics events of interest.

\begin{figure}[!ht]
\begin{center}
\includegraphics[width=0.45\columnwidth]{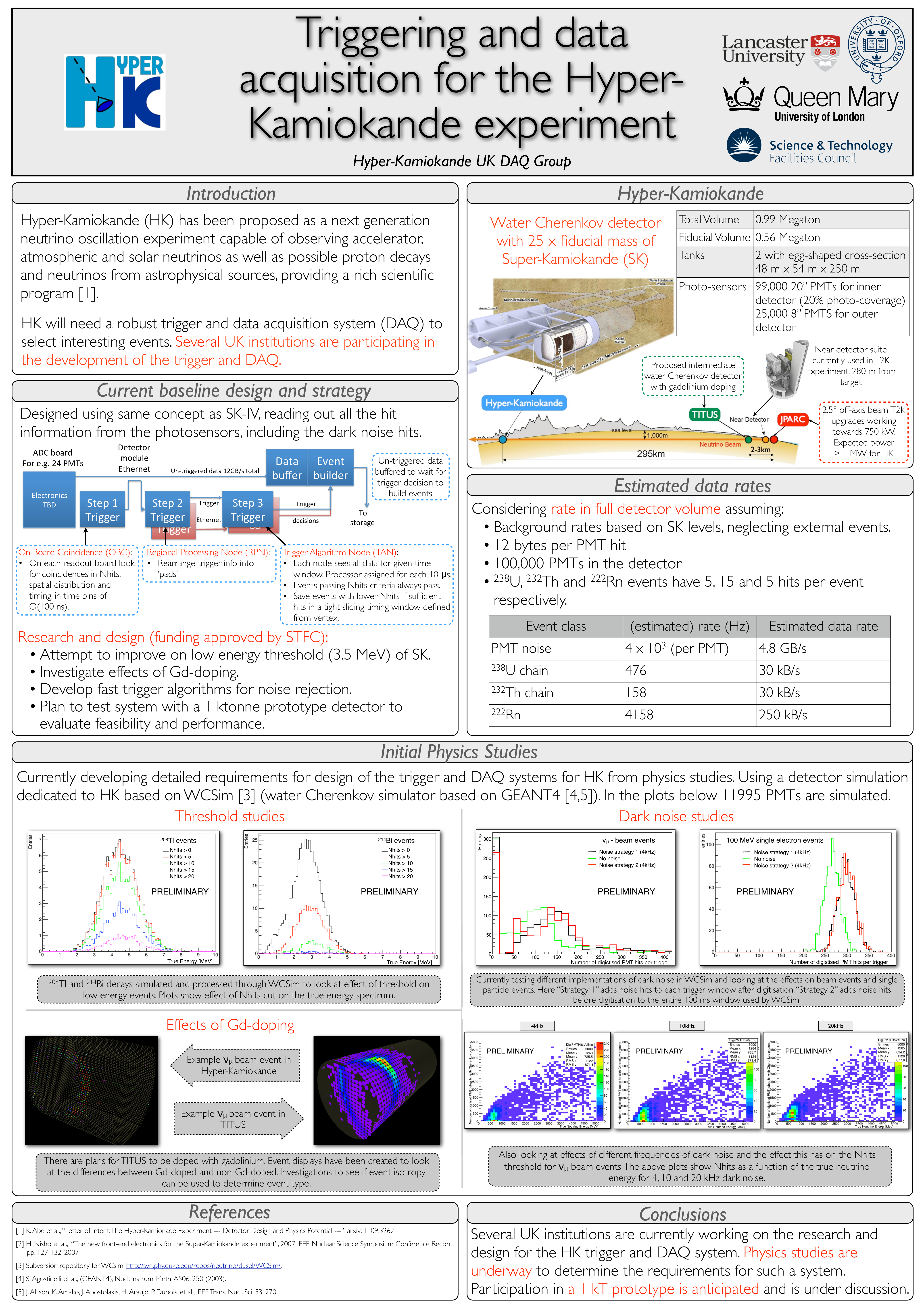}
\includegraphics[width=0.45\columnwidth]{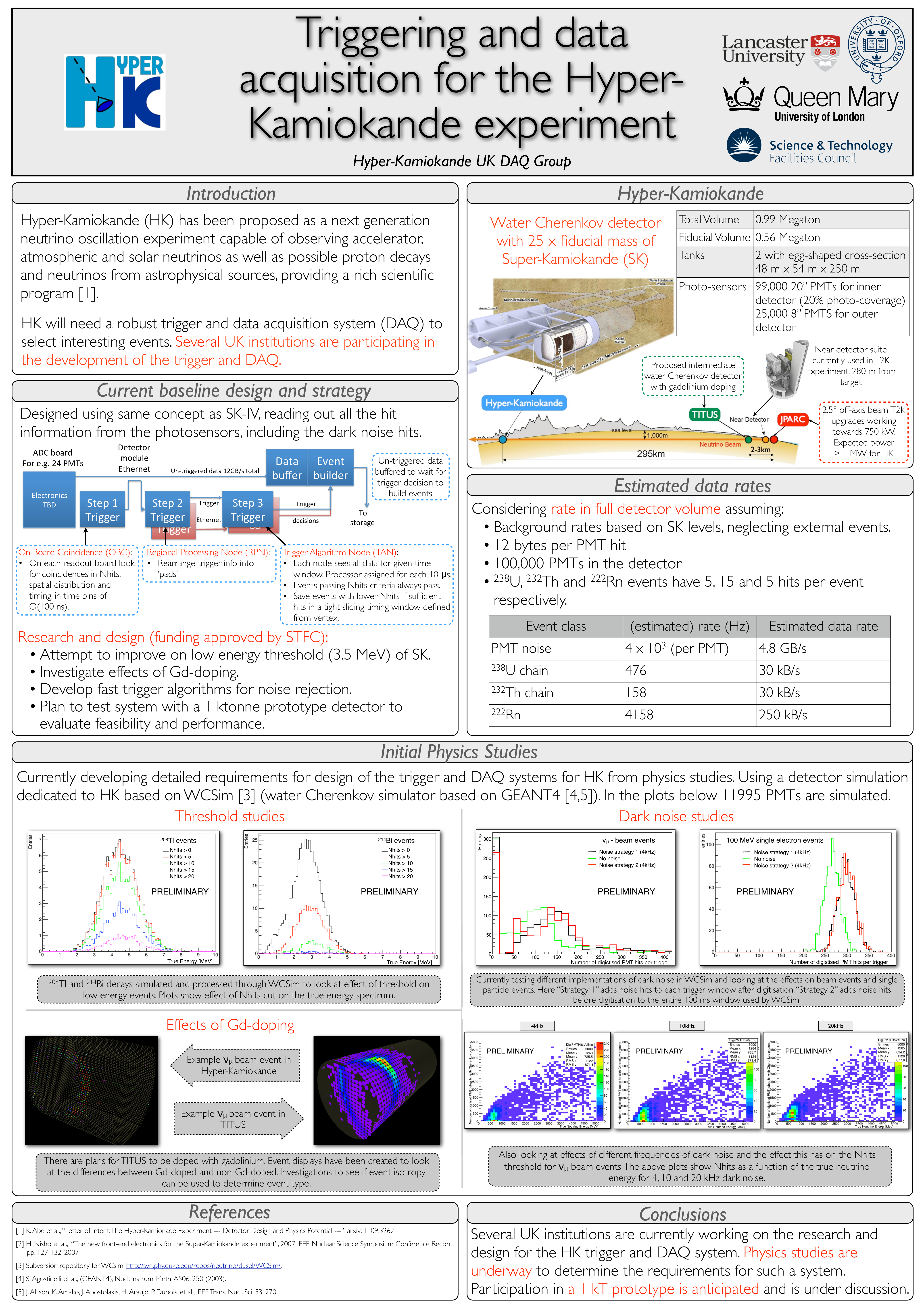}
\caption{Number of digitized PMT hits per trigger for beam events (left) and 100~MeV electrons (right) with
an Nhits trigger threshold of 25, for different implementations of the dark noise. Strategy 1 one adds hits to 
each trigger window after digitization. Strategy 2 adds noise before digitization to the entire 100~ms window.}
\label{fig:darknoise}
\end{center}
\end{figure}

\subsection{Dark noise studies}

Studies are also being performed to investigate the most realistic way to implement dark noise in WCSim. 
Figure~\ref{fig:darknoise} demonstrates two different implementations of dark noise for beam events and for 
100~MeV electron events using WCSim v1.1.1. One strategy adds dark noise hits to the entire event window before digitization, 
whereas the other strategy adds hits to each trigger window after digitization, which can be computationally
faster but limiting for trigger studies. Other implementations are currently being explored. Future work will
also focus on modelling possible correlations in dark noise hits that
can occur when the dark noise comes from radioactive decays within the PMTs.

\subsection{Gadolinium doping studies}

\begin{figure}[!ht]
\begin{center}
\includegraphics[width=0.45\columnwidth]{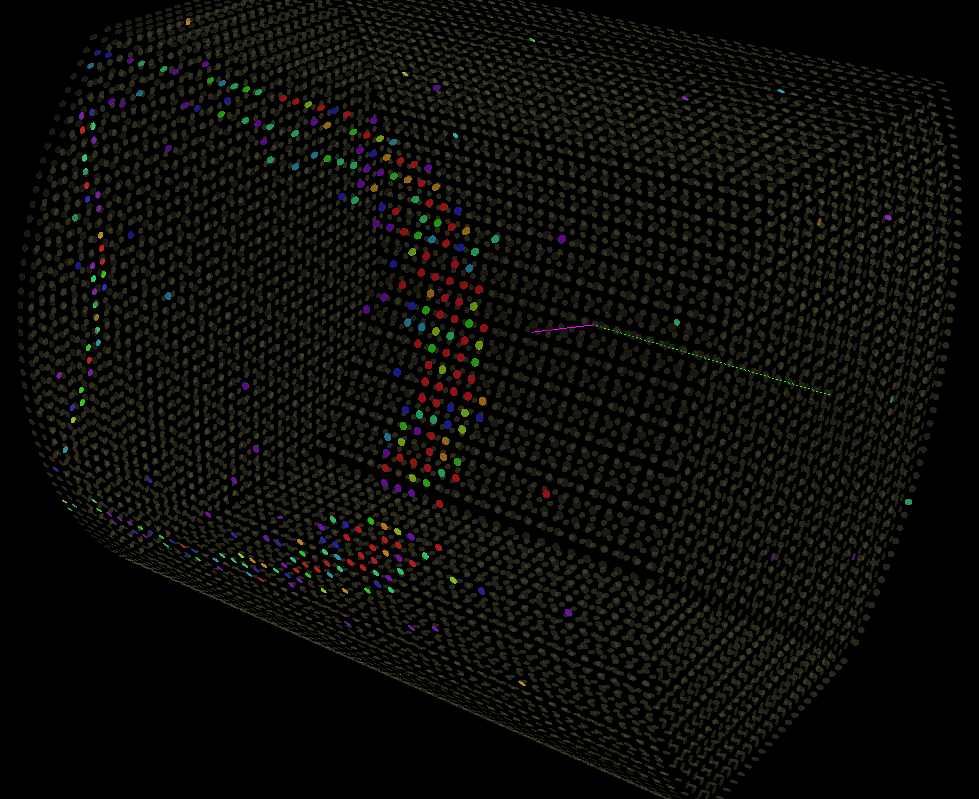}
\includegraphics[width=0.45\columnwidth]{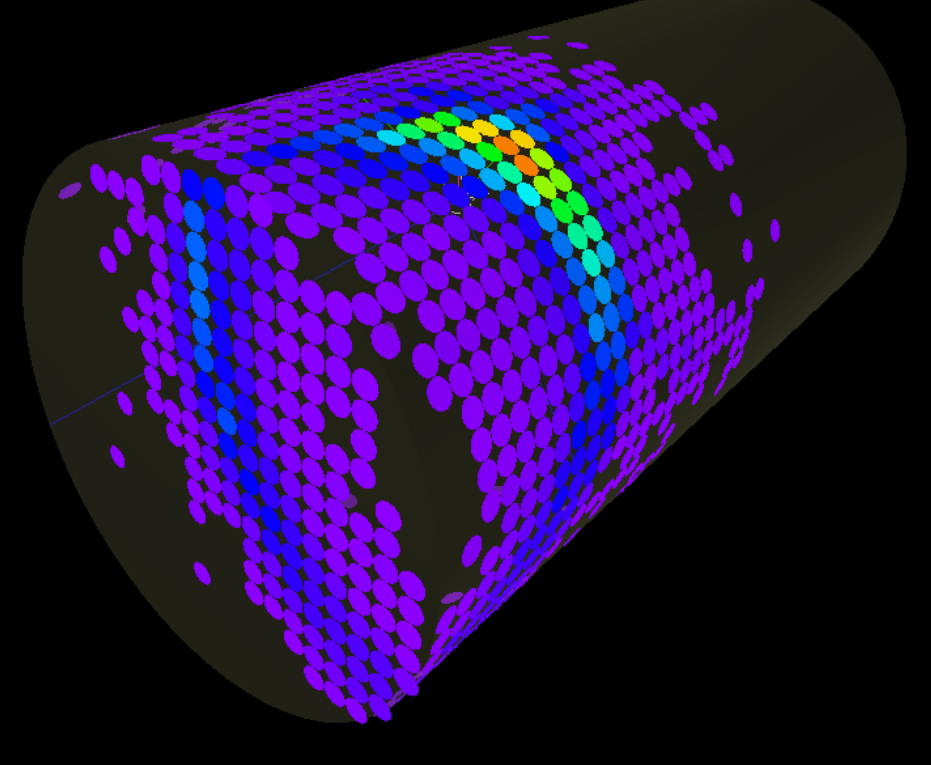}
\caption{Example event displays for HK (left) and the proposed near detector TITUS (right). Both displays
show an example $\nu_{\mu}$ beam event.}
\label{fig:eventdisplay}
\end{center}
\end{figure}

There are plans for the intermediate detector of HK to be doped with gadolinium (Gd). Event displays have been created to study
the differences between Gd-doped and non-Gd-doped events to see if event isotropy can be used to determine event type.
Figure~\ref{fig:eventdisplay} shows example event displays for $\nu_{\mu}$ beam events in HK and TITUS.

\section{Summary}

Several UK institutions are currently working on the research and design for the HK trigger and DAQ system. 
Physics studies are underway to determine the requirements for such a system. These studies will inform the
technical design of the DAQ and trigger system for HK.


\bigskip

\begin{thebibliography}{99}


\bibitem{Abe:2011ts}
  K.~Abe, T.~Abe, H.~Aihara, Y.~Fukuda, Y.~Hayato, K.~Huang, A.~K.~Ichikawa and M.~Ikeda {\it et al.},
  arXiv:1109.3262 [hep-ex].
  
\bibitem{Abe:2011ks}
  K.~Abe {\it et al.}  [T2K Collaboration],
  Nucl.\ Instrum.\ Meth.\ A {\bf 659} (2011) 106.
  
\bibitem{Yamada:2010zzc}
  S.~Yamada {\it et al.}  [Super-Kamiokande Collaboration],
  IEEE Trans.\ Nucl.\ Sci.\  {\bf 57} (2010) 428.
  
\bibitem{Agostinelli:2002hh}
  S.~Agostinelli {\it et al.}  [GEANT4 Collaboration],
  Nucl.\ Instrum.\ Meth.\ A {\bf 506} (2003) 250.
  
  \bibitem{WCSim:2015}
  Subversion repository for WCsim: http://svn.phy.duke.edu/repos/neutrino/dusel/WCSim/.
  



\end{thebibliography}
\end{document}